\newcommand{\Pl}{\partial}
\newcommand{\ts}{\textstyle}
\newcommand{\bee}{\begin{equation}}
\newcommand{\ene}{\end{equation}}
\newcommand{\beea}{\begin{eqnarray}}
\newcommand{\enea}{\end{eqnarray}}
\newcommand{\vp}{ v_{0}'}

\newcommand{\gpe}{ \nabla_{\perp}}

\newcommand{\fder}[2]{\frac{{\ts d \/ #1}}{{\ts d\/ #2}}}

\newcommand{\fpar}[2]{\frac{{\ts \Pl \/ #1}}{{\ts \Pl \/ #2}}}
\newcommand{\nder}[3]{\frac{{\ts d^{#1} \/ #2}}{{\ts d \/ #3^{#1}}}}
\newcommand{\npar}[3]{\frac{{\ts \Pl^{#1} \/ #2}}{{\ts \Pl \/ #3^{#1}}}}

\documentclass[12pt]{iopart} 
\usepackage{iopams} 
\usepackage{graphicx}

\begin{document}


\title{Viscosity gradient driven instability of `shear mode' in a strongly coupled  plasma}

\author{D. Banerjee, M. S. Janaki and N. Chakrabarti}

\address {Saha Institute of Nuclear Physics, 1/AF Bidhannagar Kolkata - 700 064, India.}

\ead{\mailto{debabrata.banerjee@saha.ac.in}}

\author{M. Chaudhuri}

\address{ Max-Planck-Institut f\"{u}r extraterrestrische Physik, 85741 Garching, Germany}

\begin{abstract}
      The influence of viscosity gradient (due to shear flow) on low
      frequency collective modes in strongly coupled dusty plasma is analyzed.
      It is shown that for a well known viscoelastic plasma model, the velocity shear
      dependent viscosity  leads to an instability of the shear mode. The inhomogeneous
      viscous force and velocity shear coupling supply the free energy for the instability.
      The combined  strength of shear flow and viscosity gradient wins over any
      stabilizing force and makes the shear mode unstable. Implication of such a
      novel instability and its applications are briefly outlined.
\end{abstract}
\maketitle
{\tableofcontents}

\section{Introduction}
\label{intro}

         In complex physical systems, for example multispecies charged
         fluid (dusty plasma), various physical processes exist and
         interact simultaneously. The stability properties of such systems
         are complicated due to the presence of various free energy sources
         which ultimately lead to instabilities. Despite  the large
         experimental and theoretical effort, it has not been possible to
         identify  all the  sources of free energies available in such a
         complex system. In this work we have identified a
         free energy source (viscosity gradient due to velocity shear) in a
         complex dusty plasma system and demonstrated a novel
         instability of the `shear mode' due to the combined effect of viscosity
         gradient and velocity shear. The collective modes in dusty plasmas have
         been the subject of serious study in recent years  due to
         their novel character and wide applications.  Normally in a
         three-component plasmas, besides the electrons and the most
         abundant ion species, there is an additional  species with a
         different mass and charge and whose abundance
         is not negligible compared to the other constituents. This additional
         heavy micron-size  species, having a wide range of values for
         the mass-to-charge ratio, is referred to as `dust' in the
         dusty-plasma literature \cite{rao}. The presence of the new
         species is expected to result in new  effects in the
         collective-mode behavior in the plasma \cite{gang}. This is because the
         various species are mutually coupled through the electro-magnetic
         forces. In such a scenario the plasma is able to support many new
         modes as compared to those in a simple electron-ion plasma\cite{pkaw}.
          Due to the large
         amount of charge on a single dust particle, the dust fluid can also
         exhibit strong coupling behavior which can show strong viscous
         properties of the medium even leading to viscoelastic behavior\cite{pkaw}.
         The strongly coupled complex plasma has been realized in different experiments\cite{jchu,haya,thom,htho}.
         The strength of the coupling is characterized by the Coulomb coupling
         parameter $\Gamma = q_d^2/(k_B T_d a) $ where $q_d$ is the charge on the
         dust grains, $ a (\simeq n_d^{-1/3})$ is the average distance between
         them for density $n_d$, $T_d$ is the temperature of the dust component
         and $k_B$ is the Boltzmann constant\cite{ikej}. In the regime of $\Gamma$ from $1$
         to $\Gamma_{c}$ (a critical value beyond which system becomes crystalline)
          both viscosity and elasticity are equally important and this property
          together is known as visco-elasticity. When $\Gamma>\Gamma_{c}$, viscosity
          disappears and  only elasticity reigns over the system.
         Experiments \cite{thom} have also shown that as $\Gamma$ increases,
         the dust components becomes strongly coupled and for large $\Gamma$ values,
         the dust component becomes crystalline. This phenomenon, the plasma condensation is
         useful in studying phase transitions \cite{htho,morf} and low
         frequency wave propagation \cite{piep,melz}.
        It has been shown that the strong
        correlations make dusty plasma system rather rigid so that it can support a
        transverse shear mode\cite{pkaw}. This `shear mode' has also been found experimentally \cite{pram} and
        its variant theoretically\cite{skla}.

	    An interesting property observed in the case of complex dusty plasmas
	    is the strong density dependence of the viscosity parameter \cite{stei} and owes its
	    existence to the large amount of charge on each dust particle.
        Recent experiments \cite{ivle} reveal that complex-plasma fluid has
        the signature of non-Newtonian property similar to other non-Newtonian fluids.
        Beyond some critical value of velocity shear rate, the medium shows shear-thinning
        property which means that the coefficient of viscosity decreases with the increase in shear
        rate. Based on experimental input,  Ivlev et. al. \cite{ivle} have shown the power law
	    dependence of viscosity on
        velocity shear. The experiment has been done  with  gas induced shear flow for different discharge
        currents and also by applying laser beams of different power.
        Hence they have measured shear viscosity for a wide range of velocity shear rates and confirmed shear
        thinning property over a considerable range. Very recently similar experiment has also been
         reported by  Gavrikov et al. \cite{jpps} in a dusty plasma liquid. simulation work in this direction has also been reported \cite{donko}.

        Motivated by these experimental results, we have  investigated the effect of equilibrium viscosity gradient
        on a shear mode in a strongly coupled plasma. Shear thinning property with generalized Oldroyd-B model \cite{arad,nada}has been studied in neutral viscoelastic fluid.
        In this work, we have demonstrated indeed that a novel instability exists due to the coupling of shear
         flow to the velocity fluctuations via velocity shear induced viscosity gradient.

         \section{Basic equations and equilibrium} \label{sec:bas}
         In a standard fluid description of dusty plasma for studying low
          frequency ($\omega \ll k v_{th e}, k v_{th i}$ where  $v_{th e,i}$  correspond to the thermal
          velocities of  electrons and ions respectively  and $k$ is a typical wave vector) phenomena normally
          we treat electrons and ions as a light fluid which can be
          modeled by a Boltzmann distribution neglecting their inertial effects in the momentum
          equations.  This is justified because
          due to higher temperature and smaller electric charge compared to dust they can easily
          thermalize and give rise to Boltzmann distribution.
          The ion and electron densities can be written this way: $n_{e(i)} = n_{0e(0i)}\exp{[\pm e \phi/T_{e(i)}]}$, where $n_{0e(0i)}, T_{e(i)}$ correspond to   the equilibrium densities and temperatures for the electrons (ions).   Here, $\phi$ is the electrostatic potential.

          The dust component on the other hand, can be described
          by generalized hydrodynamic (GH) equation described in  Frenkel's book \cite{fren}. We follow the same
          procedure and write the generalized equation of motion of dust fluid in a viscoelastic medium
        \bee
        \left(1+ \tau \fpar{}{t} \right)\left[\rho \left(\fpar{}{t}+{\bf v}\cdot \nabla\right){\bf v}
            - n_d q{\bf E} + \nabla p_d \right ] = \fpar{\sigma_{ij}}{x_{j}}
        \label{deq}
        \ene
        where  ${\bf v}$ is the dust fluid velocity, $\rho=m_d n_d$ is the mass density of dust fluid, $n_d$ is corresponding
        number density, $q$ is the  charge on a dust particle, $p_d(=n_d T_d)$ is
        the dust pressure where $T_d$
        is the dust temperature. The parameter  $\tau$ is the relaxation time of the  medium \cite{fren} and viscosity tensor $\sigma_{ij}$ is given for an incompressible medium by
        \[
        \sigma_{ij} = \eta(S) \left(\frac{\partial v_{i}}{\partial x_{j}} + \frac{\partial v_{j}}{\partial x_{i}}\right).
        \]
        Here  $\eta$ is  the coefficient of shear viscosity.  For a Newtonian fluid $\eta $ is constant. However, for a
        non-Newtonian fluid, $\eta$ depends on the scalar invariants of strain tensor.
        For an incompressible fluid, it has been shown that \cite{byro, malk} the scalar invariant can be written as
         \[
         I = \sum_i \sum_j \left(\frac{\partial v_{i}}{\partial x_{j}} + \frac{\partial v_{j}}{\partial x_{i}}\right)\left(\frac{\partial v_{i}}{\partial x_{j}} + \frac{\partial v_{j}}{\partial x_{i}}\right).
         \]
         In this work we consider an incompressible plasma with constant mass density and since the medium is non-Newtonian, the viscosity coefficient can be  considered to be a function of the scalar invariant. Hence, the viscosity parameter is taken to be of the form $\eta(S)$ where $ S= \sqrt{ I/2}$.

\begin{figure}
                    \centering
                    \includegraphics[width=2.5in,height=2.0in]{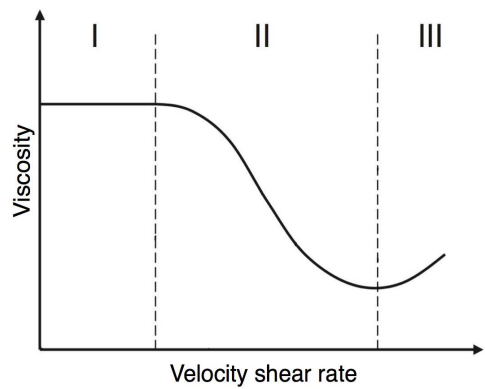}

                    \caption{Figure shows the dependence of shear viscosity with velocity shear rate where in the
                    low shear rate region it is newtonian(I), for high shear rate shear-thickening(III).
                    Our interest is to study the region-II where $\eta$ decreases with velocity shear rate.}
                    \label{sp}
                \end{figure}

         It has been shown that  in the kinetic limit $\tau \partial/\partial t \gg 1$, linearized Eq.(\ref{deq}) gives rise to a `shear wave' whose velocity is given by $V_s= \sqrt{\eta/\rho \tau}$, where $\eta$ is a constant\cite{pkaw}. We would like to
         investigate the dynamics of this mode in presence of velocity shear dependent viscosity coefficient. In the
         kinetic limit,
         $1$ can be neglected with respect to $\tau \partial/\partial t$ in the above equation(\ref{deq}).
         We assume, that the equilibrium velocity is directed along $y$ direction and has variation in $x$-direction, i.e. ${\bf v}_0 = v_{y 0}(x) \hat e_y$, where $\hat e_y$ is a unit vector along $y$ direction.
         It is clear that the left hand side of  equation (\ref{deq}) will not contribute in equilibrium
	situation, so that the equilibrium is described by the equation
        \bee
        \fder{}{x} \left[ \eta(S_0) \frac{dv_{0y}}{dx} \right]= 0.
        \label{eql}
        \ene
	where $S_0$ is the equilibrium value of the shear parameter.
        For small velocity fluctuations, we can write  $S=S_0+S_1$,
	and  it can be shown that $S_0= dv_{y0}/dx$
        and $S_1=(\partial v_{1x}/{\partial y} + \partial v_{1y}/{\partial x}).$
        Recently, Ivlev {\it et. al} \cite{ivle} have proposed a power-law model for the functional dependence of $\eta (S_0)$, and performed an experiment to show the shear thinning property of dusty plasmas. In this paper,
         they have
        shown that  $\eta$ remains constant for low shear rate and after some critical value $\eta$ decreases with increase of $S_0$. Shear thinning behavior exists for a wide range of velocity shear and for some very high shear rate $\eta$
        increases with $S_0$. A schematic sketch in fig.(\ref{sp}) represents the variation of shear viscosity with velocity shear rate.
        In the present work we concentrate in the shear thinning region (where $\eta$
        decreases with $S_0$) and use the same  model  consistent with the experiment. The  functional form can be
        written as,
        \bee
        \eta(S_0) = \bar{\eta_0} \left(\frac{S_0}{S_{c}}\right)^{- \frac{2\delta}{1+\delta }},
        \label{model}
        \ene
        here, $\delta $ is a positive exponent and $\bar \eta_0$ is constant having
        the dimension of viscosity coefficient. If we define $ {2\delta}/{(1+\delta)} = \alpha$,
        then the parameter $\alpha$
        is a positive non-zero constant $\alpha <2 $ and $S_{c}$ is of order unity. The power law model can only
        depict shear thinning region (region II)
        of fig.(\ref{sp}) but it neither can explain the constant behavior
        of viscosity in low shear region (I) nor the shear thickening property
        for very high shear rate (region III). In low shear region(I), the system behaves as a Newtonian fluid.
        In high shear region (III),  another power law model like $\eta \propto (S_0)^ {\bar \delta} $
         with a different positive exponent $\bar \delta$ describes the shear thickening behavior.
        When the equilibrium profile of $\eta(S_0)$ from Eq. (\ref{model}) is substituted in Eq. (\ref{eql}),
        we find that $dv_{y0}/dx$ is constant and hence we can write the equilibrium velocity
        as $v_{y0}(x) = \vp x,$ where $\vp$ is a constant having the dimension of frequency.
\section{Stability analysis}\label{sec:ana}
        We restrict our attention to two dimensional incompressible perturbations such that all
        variations are in the $x-y$
        plane. The incompressibility condition given by $\nabla \cdot {\bf v}=0$ is consistent with the equilibrium
        flow. Now we perturb the system  around this equilibrium flow writing  $ {\bf v} = v_{y 0}(x) \hat e_y + {\bf v}_{1}(x,y,t)$, and
        after a straightforward algebra we find that $v_{1x}$ satisfies a differential equation which is given by
        \bee
        \fpar{}{t}\left[ \fpar{}{t} +  \vp x \fpar{}{y}\right] \gpe^2 v_{1x}
        = \left(\frac{\eta_{0}}{\tau \rho} \right) \gpe^4 v_{1x} +\left( \frac{\eta'_{0} v'_{0}}{\tau \rho}\right)
        \left(\npar{2}{}{x}- \npar{2}{}{y}\right)^2 v_{1x},
        \label{master}
        \ene
         where $\gpe^2 = \partial^2/\partial x^2+\partial^2/\partial y^2$.
        We note here that if the velocity shear is absent
        in Eqs. (\ref{master}), we get back the shear mode
        whose dispersion relation can be written as $\omega^2 = (k_x^2+k_y^2) V_s^2.$

       For an inhomogeneous plasma the general solution of Eq. (\ref{master}) including viscosity gradient
       can be obtained in various ways. The traditional starting point of an investigation of
       linear plasma stability is the eigenvalue analysis in  which
       we assume the solution of the form
       $v_{1x} = v_1(x) \exp ( ik_y y -i \omega t)$, where $k_y$ is the wave vector in $y$ direction and
       $\omega$ is the frequency of the mode.  We note here that Fourier type of solutions have been considered
	   only in the $y$-direction, since inhomogeneity is present in the $x$-direction through
	   velocity shear.
       The perturbed variable $v_1$ in general will satisfy a differential equation which is given by
       \beea
        \frac{\omega_s^2}{k_y^2}\left(1+ \frac{\eta_0' \vp}{\eta_0}\right)\nder{4}{v_1}{x}+
        \left[\omega^2 - \omega k_y \vp x - 2 \omega_s^2\left(1-\frac{\eta_0' \vp}{\eta_0}\right)\right]
        \nder{2}{v_1}{x} \nonumber \\
        + \left[  k_y^2 \omega_s^2\left(1+\frac{\eta_0' \vp}{\eta_0}\right) - k_y^2 (\omega^2- \omega k_y \vp x) \right]v_1=0,
        \label{diff}
        \enea
        where $\omega_s^2 = k_y^2 \eta_0/\rho \tau$ is the frequency of the shear wave in an inhomogeneous plasma.
        From the above differential equation first we can do the local analysis in which one uses
       the approximation $k L \gg 1$. This implies that the
       perturbation wave length $k^{-1}$ is much smaller than the inhomogeneity scale length
       $L (=v_0/\vp)$. For the present problem, the local analysis
        is carried out by considering that the perturbed
       quantity $v_{1}$ has also an exponential variation in  $x$ i.e. $v_1 \sim \exp (i k_x x)$,
       where $k_x$ is  the wave vector in $x$ direction. Substituting in Eq. (\ref{diff})
        the dispersion equation  is obtained as
        \bee
         \omega^2 = \frac{\omega_s^2 (k_x^2+ k_y^2)}{k_y^2}
         \left[1+\frac{\eta_0'\vp }{\eta_0}
         \left(\frac{k_y^2-k_x^2}{k_y^2+k_x^2}\right)^2\right],
         \label{nld}
         \ene
        where $\eta_0'= d\eta_0/d\vp$ and $\omega \gg k_y v_0$. Now recalling the form of viscosity $\eta_0$ from Eq. (\ref{model}) we can write
        $\eta_0' \vp/\eta_0= - \alpha$. It is clear that the shear mode will be unstable
        if $\alpha > (k_x^2+k_y^2)^2/(k_y^2-k_x^2)^2$ and the growth rate is of the order of
        shear frequency.

         Next we consider the  nonlocal analysis of Eq. (\ref{diff}) in which this eigenvalue
         equation may be solved to get well behaved solutions corresponding to unstable eigenvalues.
         Here we are looking for long radial ($x$) scale solution for the differential equation and therefore
         the fourth order derivative is subdominant compared with the second. Ignoring the fourth derivative
          in Eq. (\ref{diff}), we can reproduce the character of the mode with a very little change
          (this is apparent from the dispersion relation).  This assumption
         simplifies the algebra without taking away the essential physics.
          The desired eigenvalue equation can be written as
         \bee
         \nder{2}{v_1}{x}- k_y^2 \left[\frac{\omega^2 - \omega_s^2 (1-\alpha)-\omega k_y \vp x}
         {\omega^2 - 2\omega_s^2 (1+\alpha)-\omega k_y \vp x} \right]v_1=0.
         \ene
         For the condition
         $\omega k_y v_0'/[\omega^2-2 \omega_s^2(1+\alpha)]\ll 1$, which implies that, when the shear rate is small compared
         to the frequency of the mode, the above equation can be written in terms of  the well known Weber equation which is given by
         \bee
          \nder{2}{v_1}{\xi}-(\xi^2-K)v_1=0
          \label{wb}
         \ene
         where
         \[ \xi = \left[k_y^2 \beta_2^2 \left(\frac{\beta_2-\beta_1}{\beta_1}\right)\right]^{1/4}
         \left(x+\frac{1}{2 \beta_2}\right),\;\;
         K= - \frac{k_y}{4}\left(\frac{1}{\beta_2}+\frac{3}{\beta_1}\right)\sqrt{\frac{\beta_1}{\beta_2-\beta_1}}.\]
         and
         \[ \beta_1 = \frac{k_y \vp \omega}{\omega^2 - \omega_s^2(1-\alpha)},\;\;
         \beta_2 = \frac{k_y \vp \omega}{\omega^2 - 2\omega_s^2(1+\alpha)}\]
         The solution of Eq. (\ref{wb}) for the lowest order eigenmode is given by
         \bee
          v_{1} \sim \exp \left[- \frac{1}{2}k_y \beta_2 \sqrt {\left(\frac{\beta_2-\beta_1}{\beta_1}\right)}
          \left(x+\frac{1}{2 \beta_2}\right)^2 \right]
         \ene
          representing the existence of an unstable eigenmode. The condition for bounded solution is
          $Re \left( \beta_2 [{(\beta_2/\beta_1)-1}]^{1/2} \right) > 0$. The behaviour of the eigenfunction $v_{1}$
          at $x\rightarrow \pm \infty$ is bounded and the typical mode width $\bigtriangleup \sim \left[ \frac{1}{k_y \beta_2} \sqrt{\frac{\beta_1}{\beta_2 -
           \beta_1}} \right]^{1/2}$. The corresponding dispersion relation is given by

\begin{figure}
                \centering
                \includegraphics[width=4.0in,height=4.0in]{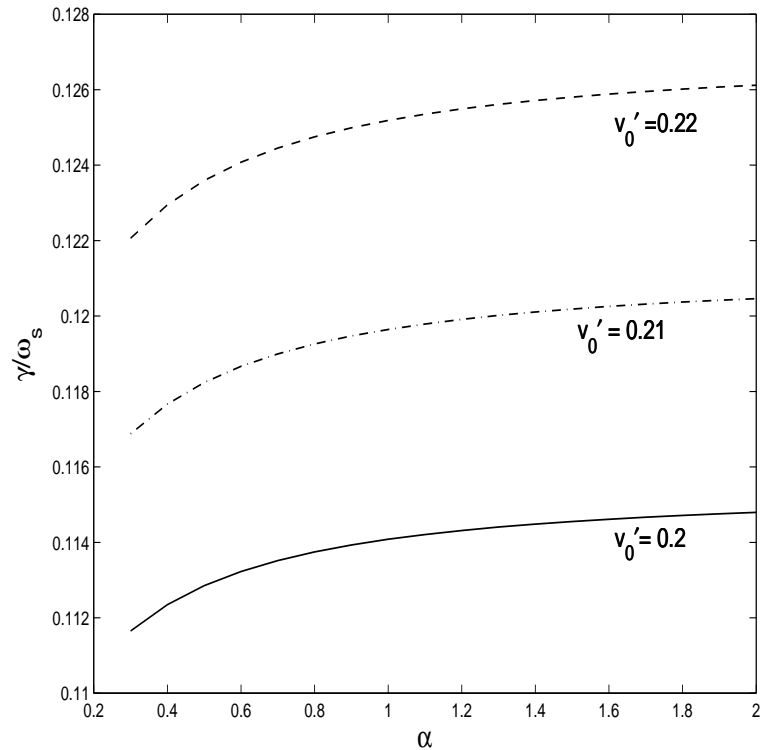}
                \caption{Normalized growth rate $\gamma/\omega_s$ as a function of
                $\alpha=|\eta_0' v_0'/\eta_0|$, shows the growth
                rate of shear mode for different velocity shear parameter}
                \label{fig:gr}
             \end{figure}

          \bee
           \omega^2- \frac{\omega_s^2}{4}(5 - \alpha)=-\vp \omega \omega_s
           \sqrt{\frac{1+3 \alpha }{\omega^2- 2 \omega_s^2(1+\alpha)}},
           \label{dr}
          \ene
          where $\alpha = -\eta_0' \vp/\eta_0 >1$ and $\omega_s^2= k_y^2 \eta_0/\rho \tau.$ In a homogeneous plasma
          i.e. when $\vp= \alpha =0$  we get back  the shear mode i.e. $\omega^2 \sim \omega_s^2$.
          In presence of velocity shear and velocity shear induced viscosity gradient we have solved Eq. (\ref{dr})
          and found that for $\alpha <2$ there is one unstable root for real $\omega>0$.
          The growth rate for the instability for the given range of $\alpha$ can be seen in the figure~\ref{fig:gr}.
          The shear mode is more unstable when velocity shear is stronger.
           \section{Summary}\label{sec:sum} %
           We have studied the effect of velocity shear induced viscosity gradient  of low frequency
          shear  waves in a viscoelastic dusty plasma. The dust dynamics
           has been modeled by including velocity shear dependent viscosity  which is the main
           ingredient to drive a new instability in a complex  plasma. The principal
           effect on the generation of the novel instability is the velocity dependence
           of viscosity that leads to a coupling between velocity fluctuations and
           equilibrium flow. The variation in the velocity is responsible for
           viscosity modulation which provides a feedback to the velocity
           through momentum equation. For a positive feedback of the
           velocity, an instability is triggered. This novel low frequency instability disappears
             when viscosity is uniform and we are left with a shear wave. We
           would like to point out that the  instability theoretically
           investigated in this work has not been observed in real experiment as
           yet; its detailed experimental investigation is therefore of
           great interest. It would be of interest therefore to look for
           shear wave driven instabilities discussed in this model
           calculation which is based upon the experimental finding
           of a shear thinning region.
        \section{References} \label{sec:bib}

        \end{document}